\documentclass[12pt]{article}

\usepackage{amsmath}
\usepackage{amssymb}
\usepackage{hyperref}
\usepackage{pict2e}        

\usepackage{framed}

\usepackage{hyperref}
\usepackage{breakurl}
\usepackage{amsmath}
\usepackage{amsfonts}
\usepackage{amssymb}
\usepackage{algorithm}
\usepackage{color}

\usepackage{pict2e}        
\usepackage[pdftex]{graphicx}
\DeclareGraphicsExtensions{.pdf,.jpeg,.png}

\newcommand{\comment}[1]{}

\newcommand{\floor}[1]{\lfloor #1 \rfloor}

\def\denseformat{
\setlength{\textheight}{9.2in}
\setlength{\textwidth}{7.1in}
\setlength{\evensidemargin}{-0.2in}
\setlength{\oddsidemargin}{-0.2in}
\setlength{\headsep}{10pt}
\setlength{\topmargin}{-0.3in}
\setlength{\columnsep}{0.375in}
\setlength{\itemsep}{0pt}
\renewcommand{\baselinestretch}{0.99}
}
\def\midformat{
\setlength{\textheight}{8.9in}
\setlength{\textwidth}{6.7in}
\setlength{\evensidemargin}{-0.19in}
\setlength{\oddsidemargin}{-0.19in}
\setlength{\headheight}{0in}
\setlength{\headsep}{10pt}
\setlength{\topsep}{0in}
\setlength{\topmargin}{0.0in}
\setlength{\itemsep}{0in}       
\renewcommand{\baselinestretch}{1.1}
\parskip=0.070in
}
\def\spacyformat{
\setlength{\textheight}{8.8in}
\setlength{\textwidth}{6.5in}
\setlength{\evensidemargin}{-0.18in}
\setlength{\oddsidemargin}{-0.18in}
\setlength{\headheight}{0in}
\setlength{\headsep}{10pt}
\setlength{\topsep}{0in}
\setlength{\topmargin}{0.0in}
\setlength{\itemsep}{0in}      
\renewcommand{\baselinestretch}{1.2}
\parskip=0.080in
}

\def\thisformat{
\setlength{\textheight}{8.8in}
\setlength{\textwidth}{5.5in}
\setlength{\evensidemargin}{0.32in}
\setlength{\oddsidemargin}{0.32in}
\setlength{\headheight}{0in}
\setlength{\headsep}{10pt}
\setlength{\topsep}{0in}
\setlength{\topmargin}{0.0in}
\setlength{\itemsep}{0in}      
\renewcommand{\baselinestretch}{1.2}
\parskip=0.080in
}

\midformat
\spacyformat
\denseformat
\thisformat

\newcommand{\authornote}[1]{}
\newcommand{\commentOut}[1]{}
\newcommand{\beq}{\begin{equation}}
\newcommand{\eeq}{\end{equation}}
\def\EE{\hbox{I\kern-.1667em\hbox{E}}}

\newcommand{\CA}{\textsc{ClipAudit}}

\begin{document}

\bibliographystyle{plain} 

\title{\CA{}: A Simple Risk-Limiting Post-Election Audit}

\author{Ronald L. Rivest\\
  MIT CSAIL \\
  {\url{rivest@mit.edu}}
}

\date{\today}
\maketitle

\begin{abstract}
  We propose a simple risk-limiting audit for elections, \CA{}.
  To determine whether candidate A (the reported winner) actually
  beat candidate B in a plurality
  election, 
  \CA{} draws ballots at random, without replacement, until
  either all cast ballots have been drawn, or until
  \[
        a - b \ge \beta \sqrt{a+b}
  \]
  where $a$ is the number of ballots in the sample for the reported winner A, and
  $b$ is the number of ballots in the sample for opponent B, and
  where $\beta$ is a constant determined a priori as a function
  of the number $n$ of ballots cast and the risk-limit $\alpha$.

  \CA{} doesn't depend on the unofficial margin (as does Bravo).

  We show how to extend \CA{} to contests with multiple winners
  or losers, or to multiple contests.

\end{abstract}  

  
\section{Introduction and Motivation}
\label{sec:introduction}

The paper is organized as follows:
Section~\ref{sec:preliminaries} provides some terminology, notation,
and orientation.  
Then
Section~\ref{sec:clipaudit} provides an overview of the \CA{}
audit method.
Section~\ref{sec:computing-betas} discusses the computation of
the key constant $\beta$.
Section~\ref{sec:expected-final-sample-size} analyzes the final
expected audit sample size.
Some \CA{} variants are presented in Section~\ref{sec:variants}.
Section~\ref{sec:discussion} provides more discussion of \CA{}.
Finally, related work is described in Section~\ref{sec:related-work}.

\section{Preliminaries}
\label{sec:preliminaries}

\paragraph{Outcomes, Ballots, Profiles.}
We assume an election designed to produce an
\textbf{outcome} from a set $\mathcal{C}$ of $C$ alternatives (or candidates).

Each of $n$ voters casts a single ballot.
A ballot specifies a single candidate, chosen by the voter.

We denote the \textbf{profile} of cast ballots as
\[
    P = \{v_1, v_2, \ldots, v_n\}\ ,
\]
where each $v_i \in \mathcal{C}$.
(One may wish to allow a cast vote $v_i$ to represent an ``overvote''
or an ``undervote'' as well, although the election outcome will never
be ``overvote'' or ``undervote.'')
The profile may be viewed as a sequence or as a multiset, since
profiles may contain repeated items (identical ballots).

\paragraph{Plurality Election.}
We assume a \emph{plurality election}: the candidate with the most votes is
the winner.  (There may be ties, which can be resolved according to
relevant election law.)

\paragraph{Post-election audits.}
We assume here that voters have cast votes on \textbf{paper ballots}
on which their choices were recorded and verified by the voter.
These paper ballots were scanned, and the electronic versions of the
ballots aggregated to provide the initial or \textbf{reported outcome}
for the election.

Confidence in the reported election outcome can be derived from a
\textbf{post-election audit}.

The paper ballots are the ``ground truth'' for the election; a
full and correct count of the paper ballots should give (essentially
by definition) the \textbf{actual (or true) outcome} for the
election.  

A ``compliance audit'' can provide assurance that the paper trail has
the necessary integrity.  For details, see Benaloh et al.~\cite{Benaloh-2011-SOBA},
Lindeman and Stark~\cite{Lindeman-2012-gentle}, and Stark and
Wagner~\cite{Stark-2012-evidence-based}.

\paragraph{Statistical post-election audits}
Instead of using a recount, it is usually more efficient to audit
using a statistical method based on hand examination of a \textbf{sample} of
the paper ballots, a method first proposed by
Johnson~\cite{Johnson-2004-election-certification}.
Such a \textbf{statistical (post-election) audit}
typically provides statistical assurance that the reported outcome is
indeed equal to the actual outcome, while examining only a relatively
small sample of the paper ballots.  In the presence of errors or fraud
sufficient to make the reported outcome incorrect, the audit may need
to examine many ballots, or even all ballots, before concluding that
the reported outcome was incorrect.

\paragraph{Risk-limiting audits.}
Stark~\cite{Stark-2008-conservative}
has provided a
refined notion of a statistical audit---that of a
\textbf{risk-limiting (post-election) audit} (or \textbf{RLA}).  If
the reported outcome is incorrect, the RLA is guaranteed to have a
large, pre-specified chance $(1-\alpha)$ of 
examining all cast ballots and thereby correcting the reported outcome.
In other words, if the reported outcome is incorrect, the audit will
accept the reported outcome as correct with probability at most~$\alpha$.

Here $\alpha$ is an audit parameter (for example, $\alpha=0.05$).

Lindeman and Stark have provided a ``gentle introduction''
to RLAs~\cite{Lindeman-2012-gentle}.  General overviews of post-election
audits are available from Lindeman et al.~\cite{Lindeman-2008-principles},
Norden et al.~\cite{Norden-2007-post-election-audits}, 
and the Risk-Limiting Audit Working Group~\cite{Bretschneider-2012-risk}.
Stark and Wagner~\cite{Stark-2012-evidence-based} promulgate the notion
of an ``evidence-based election,'' which includes not only a risk-limiting
audit but also the larger goals of ensuring that the evidence trail has
integrity.

A variety of statistical methods for providing RLAs have been
developed~\cite{%
  Stark-2008-sharper,%
  Hall-09-implementing-rlas,%
  Stark-2009-cast,%
  Stark-2009-efficient,%
  Stark-2009-risk-limiting,%
  Stark-2010-rla-cluster-size,%
  Stark-2010-super-simple,%
  Checkoway-2010-single-ballot,%
  CaliforniaSOS-2011-post-election-rla,%
  Lindeman-2012-bravo,%
  Sarwate-2013-risk-limiting,%
  Stark-2014-verifiable-elections%
}.
We also note the availability of online tools for risk-limiting
audits~\cite{Stark-2015-tools}.
We compare our \CA{} proposal with
Bravo~\cite{Lindeman-2012-bravo}, one of the best and best-known
RLA methods.

\section{\CA{} overview}
\label{sec:clipaudit}

\CA{} is, like most statistical post-election audits,
structured as a sequential decision-making procedure.

Randomly chosen ballots are examined one at a time, until
a \emph{stopping rule} says that the audit should stop (accepting
the reported outcome as correct), or until all $n$ ballots
in the profile have been examined (thereby revealing the true
outcome, which may or may not be equal to the reported outcome).

\CA{} is a \emph{ballot-polling audit}: it looks only at the
paper ballots, and does not consider the electronic version
of the ballot data (cast vote record, or CVR)
produced by the initial machine-scan of the
paper ballots.  (A \emph{comparison audit} would do that, but
\CA{} is not a comparison audit. Comparison audits are generally
more efficient than ballot-polling audits, and are thus to be
preferred to ballot-polling audits when CVRs are available.)

With \CA{}, ballots are sampled \textbf{without replacement}.
(A variant of \CA{} can easily be defined that use sampling
with replacement, but such an approach is somewhat less
efficient.)

Let $\alpha$ be the desired risk-limit.

\CA{} should accept an
incorrect reported outcome as correct with probability at most $\alpha$.

If the reported outcome is incorrect, \CA{} will, with probability at
least $1-\alpha$, proceed to determine the correct outcome by
examining \emph{all} of the paper ballots, in accordance with the
definition of a risk-limiting audit.

We first describe \CA{} for the simple case that there are only two
candidates.  Its extension to handle multiple candidates is described
in Section~\ref{sec:multiple-candidates}.

Let the two candidates be $A$ and $B$.
In the ballots sampled by the audit so far,
let $a$ denote the number of votes seen for $A$
and let $b$ denote the number of votes seen for $B$.

Let $\beta = \beta(n, \alpha)$ be a constant depending on the number
$n$ of cast ballots and the desired risk limit $\alpha$.  To get
started, the reader may imagine that $\beta=3$ 
(which is about right for many typical values of $n$ and $\alpha=0.05$);
more accurate values can be obtained from Figure~\ref{fig:beta-chart}.

\CA{} has a very simple structure:

\medskip
\fbox{%
\begin{minipage}{0.90\textwidth}
\noindent {\bf \CA{} Procedure:}
\begin{itemize}
\item Determine $\beta = \beta(n,\alpha)$ from
  Figure~\ref{fig:beta-chart}, rounding $n$ up and $\alpha$ down
  if necessary to find a relevant table entry.
\item
Draw cast ballots at random without
replacement, keeping track of the number $a$
of votes seen for the reported winner $A$ and
the number $b$ of votes seen for the opposing candidate, until either
\begin{equation}
       a - b > \beta \sqrt{a+b}\ ,
\label{eqn:stopping-condition}
\end{equation}
in which case the reported winner $A$ is
accepted as the correct outcome, or until
all cast ballots are examined, in which case
the correct election outcome is revealed.
\end{itemize}
\end{minipage}
}

\section{Computing $\beta$}
\label{sec:computing-betas}

\comment{
\begin{figure}
\begin{center}
\begin{tabular}{c|c|c|c|c|c|c}
  $n\backslash \alpha$   & 0.010 &  0.020 &  0.050 &  0.100 &  0.200 &  0.500 \\ \hline
     100 & 2.673 &  2.480 &  2.236 &  2.000 &  1.732 &  1.147 \\ 
     300 & 2.869 &  2.711 &  2.449 &  2.183 &  1.890 &  1.347 \\ 
    1000 & 3.104 &  2.898 &  2.586 &  2.309 &  2.000 &  1.414 \\ 
    3000 & 3.190 &  2.994 &  2.646 &  2.384 &  2.091 &  1.500 \\ 
   10000 & 3.297 &  3.053 &  2.760 &  2.492 &  2.183 &  1.635 \\ 
   30000 & 3.379 &  3.138 &  2.828 &  2.592 &  2.254 &  1.722 \\ 
  100000 & 3.418 &  3.244 &  2.916 &  2.646 &  2.325 &  1.757 \\ 
  300000 & 3.462 &  3.262 &  2.967 &  2.673 &  2.380 &  1.815 \\ 
 1000000 & 3.500 &  3.299 &  2.992 &  2.742 &  2.433 &  1.883 \\ 
 3000000 & 3.540 &  3.331 &  3.045 &  2.782 &  2.475 &  1.933 \\ 
\end{tabular}
\end{center}
\caption{(OLD)Approximate values of $\beta(n,\alpha)$ for various values of $n$ and
  $\alpha$, computed using simulation; 
  number of trials per entry is 10,000.}
\end{figure}
} 

\begin{figure}
\begin{center}
\begin{tabular}{c|c|c|c|c|c|c}
  $n\backslash \alpha$   & 0.010 &  0.020 &  0.050 &  0.100 &  0.200 &  0.500 \\ \hline
     100 & 2.683 &  2.500 &  2.236 &  2.000 &  1.732 &  1.155 \\ 
     300 & 2.887 &  2.694 &  2.425 &  2.145 &  1.877 &  1.343 \\ 
    1000 & 3.054 &  2.864 &  2.546 &  2.294 &  2.000 &  1.414 \\ 
    3000 & 3.184 &  3.000 &  2.670 &  2.401 &  2.095 &  1.511 \\ 
   10000 & 3.290 &  3.077 &  2.770 &  2.496 &  2.183 &  1.633 \\ 
   30000 & 3.357 &  3.144 &  2.828 &  2.556 &  2.240 &  1.715 \\ 
  100000 & 3.411 &  3.206 &  2.889 &  2.638 &  2.324 &  1.747 \\ 
  300000 & 3.487 &  3.273 &  2.958 &  2.684 &  2.375 &  1.817 \\ 
 1000000 & 3.530 &  3.309 &  3.000 &  2.734 &  2.438 &  1.890 \\ 
 3000000 & 3.560 &  3.352 &  3.040 &  2.782 &  2.474 &  1.937 \\ 
\end{tabular}
\end{center}
\caption{Approximate values of $\beta(n,\alpha)$ for various values of $n$ and
  $\alpha$, computed using simulation; 
  number of trials per entry is 1,000,000.
  The computation of these values took about 60 hours on a MacPro laptop, using
  a short Python3 program.
}
\label{fig:beta-chart}
\end{figure}

The \CA{} procedure depends on a well-defined value $\beta=\beta(n,\alpha)$,
which depends upon the number $n$ of cast votes and upon the
risk limit $\alpha$.

When using \CA{} to audit a particular election,
one can determine an appropriate value of $\beta$ use
by consulting Figure~\ref{fig:beta-chart}.  If the exact values
of $n$ and $\alpha$ for the election audit are not available in
the table, one can round $n$ up and round $\alpha$ down as needed.

In this section we describe how to compute $\beta(n,\alpha)$.

\subsection{Computation of $\beta$ using simulation}

Let $n$ be an even integer.

Let
\[
     X \in \{-1, 1\}^n  
\]
be a randomly chosen vector \textbf{with zero sum}, 
and let 
\begin{equation}
      S_t = X_1 + X_2 + \cdots + X_t
\label{eqn:random-walk}
\end{equation}
for $1\le t \le n$
be the associated random walk.  $S_t$ corresponds
to the margin of candidate A over B within a sample
of size $t$, as the sample size $t$ increases, when A and
B are in fact \emph{tied} in the overall profile of size~$n$.

Note that sampling is done from the profile ``without replacement,''
so that at most $n$ ballots can be drawn for the sample, and when all
$n$ are drawn, the two candidates are tied (since $S_n=0$).

Let $S'_t$ be the observed margin in the actual audit for the
sample of size $t$.  Assume that A is the announced
winner, and that $S'_t$ is the margin of A over B in the
sample.  

The stopping condition for \CA{} is
\begin{equation}
     S'_t > \beta \sqrt{t}
\label{eqn:stopping}
\end{equation}
where $\beta$ is chosen so that a random walk 
\begin{equation}
     S = (S_1, S_2, ..., S_n)
\end{equation}
for a tied race has chance exactly $\alpha$ of 
stopping (i.e. satisfying (\ref{eqn:stopping})
for some $t$, $1\le t \le n$).

\CA{} is thus a risk-limiting audit (RLA) since if Bob
really won, the chance that the audit stops and accepts Alice is not
greater than the chance that the audit stops and accepts Alice if 
there is actually a tie.

Computing $\beta$ can be done in a couple of ways: simulations or
approximate computations via dynamic programming.  (``Approximate''
because this method assume that the sampling is with replacement,
rather than without replacement.)  An analytic approach might also
work, but I haven't seen how to do that yet.

In any case, the simulation method is the fastest.

Computations yield, e.g.
\[
    \beta(n=10000, \alpha=0.05) \approx 2.77\ .
\]
Thus, for a 10000-vote election with announced winner Alice
with $\alpha=0.05$ the audit would stop when the margin 
$a-b$ for Alice over Bob exceeds 
\[
    2.77 \sqrt{a+b}
\]
when the sample shows $a$ votes for Alice and $b$ votes for Bob.

\paragraph{Simulation method}
We now describe the computation of $\beta(n,\alpha)$ by the
simulation method in more detail.
Let $T$ denote the desired number of trials (simulation runs).
We use $T=10^6$ in our computations.

For each trial $i$, $1\le i\le T$:
\begin{enumerate}
\item Generate a random vector $X$ of length $n$ having zero
  sum.  Each entry in $X$ is $+1$ or $-1$. (If $n$ is odd,
  then let $X$ have sum $+1$.)
\item Compute the associated random walk via
  equation~(\ref{eqn:random-walk}).
\item Compute
  \begin{equation}
    \beta_i = \max_{1\le t\le n} \frac{S_t}{\sqrt{t}}\ .
  \end{equation}
\item  
Return $\beta(n, \alpha)$ as that $k$th smallest value $\beta_i$,
where $k=\floor{(1-\alpha)T}$.  Here ``$k$th smallest'' means the value
that is larger than or equal to $k$ values in the set---that is, the
$k$th element if the values in the set are sorted into increasing order.
\end{enumerate}

\paragraph{Formula for estimating $\beta(n,\alpha)$}

We worked to fit a formula to the entries in Figure~\ref{fig:beta-chart}.
Our best fit was to the formula:
\begin{equation}
    \beta(n,\alpha) \approx 0.075 \ln(n) + 0.700\, \textrm{isf}(\alpha) + 0.860\ ;
\label{eqn:best-fit-for-beta}
\end{equation}
where $\textrm{isf}(\alpha)$ is the inverse survival function for
the standard normal distribution---the value $x$ such that
$\Phi(x) = 1 - \alpha$, and $\Phi$ is the cdf for the standard
normal distribution.
Over the values shown in the table the approximation was always within
0.16, and typically less than 0.05.  

To get a formula that is a good fit but always an upper bound on the
desired $\beta$ value, it suffices to raise the constant in the
formula to 1.00:
\begin{equation}
    \beta(n,\alpha) \le 0.075 \ln(n) + 0.700\, \textrm{isf}(\alpha) + 1.000\ ;
\label{eqn:best-fit-for-beta-upper-bound}
\end{equation}

The \CA{} user may find it convenient
to derive $\beta(n,\alpha)$ using
formula~(\ref{eqn:best-fit-for-beta}) 
or formula~(\ref{eqn:best-fit-for-beta-upper-bound})
rather than using the table.

\section{Expected Final Sample Size}
\label{sec:expected-final-sample-size}

Let $m$ denote the actual fractional margin in favor of $A$---the
fraction of votes for $A$ minus the fraction of votes for $B$.
If the actual fractional margin in favor of $A$ is $m$, then the expected
value of $S_t$ is $mt$.

Thus, $S_t\ge \beta \sqrt{t}$ is expected to occur when
\begin{equation}
    t \ge \frac{\beta^2}{m^2}\ .
\label{eqn:clipaudit-sample-size}
\end{equation}

Compare with Bravo result:
equation (7) in \cite{Lindeman-2012-bravo}
says that Bravo should terminate once
\begin{equation}
    t \ge \frac{2 \ln(1/\alpha)}{m^2}\ .
\label{eqn:bravo-sample-size}
\end{equation}

Comparing \CA{}'s $\beta^2$ with Bravo's $2\ln(1/\alpha)\approx 6$,
(for $\alpha=0.05$), 
we see that Bravo has an advantange for $\beta > \sqrt{6} = 2.449$,
\emph{assuming that the reported margin is correct}.

\paragraph{Example}

Consider auditing an election with $n=50000$ cast ballots, of which
60\% are reported for candidate A and 40\% are reported for candidate
B, with a risk limit of $\alpha=0.10$.

For Bravo, formula~(\ref{eqn:bravo-sample-size}) gives an estimated
sample size of 115; experimental results (shown in Table 1 of
Lindeman et al.~\cite{Lindeman-2012-bravo}) show an average sample
size (ASN) of 119, assuming that the reported fraction of votes for
each candidate is correct.

For \CA{}, we use the value $\beta=2.568$ from
formula~(\ref{eqn:best-fit-for-beta}); then
formula~(\ref{eqn:clipaudit-sample-size}) gives an estimated sample
size of 165; experimental results show an average sample size of 143.

We see that (at least for this example), using \CA{} rather than Bravo
incurs a performance penalty (in terms of number of ballots examined)
on the order of 20\%.

However, one advantage of \CA{} is that its performance is
\emph{insensitive} to the reported vote fractions.  If the reported
vote fractions had been 70\% for candidate A and 30\% for candidate B
(while the true fractions remained at 60\% and 40\%), Bravo would have
performed a \emph{full recount} with high probability, examining all
50000 ballots, whereas the expected workload by \CA{} would be
unchanged.  Intuitively, Bravo is trying to see if the true vote
fractions are ``closer'' to the reported vote fractions or to vote
fractions representing a tied outcome; if the true vote fractions are
``closer'' to a tie (even if they support the reported outcome), then
the Bravo audit will typically be forced into a doing a full recount.

Thus, one might adopt \CA{} in situations where the reported vote
fractions are a bit suspect, and/or where there is a desire to avoid
the possibility of a full recount being caused by incorrect reported
vote fractions.

\section{Variants}
\label{sec:variants}
\label{sec:multiple-candidates}


\paragraph{Multiple candidates.}
The \CA{} method can be easily extended to handle
more than two candidates, possibly with multiple winners.
Suppose the election has $C$ candidates, and that the
top $W$ vote-getters will be declared the \emph{winners},
and that the remaining $L=C-W$ will be declared \emph{losers}.
Here either $W$ or $L$ or both may be larger than $1$.

Let $A_1, \ldots, A_W$ denote the $W$ reported winners, and
let $B_1, \ldots, B_L$ denote the $L$ reported losers.

We simply modify \CA{} to concurrently audit that each
$A_i$ defeats $B_j$, for $1\le i\le W$ and $1\le j\le L$.
Each subaudit uses the same sequence of audited ballots, paying
attention only to those ballots relevant to its task.
Each audit uses the same overall risk limit $\alpha$.
The overall audit stops and accepts the reported outcomes
if and only if all of the subaudits have stopped and accepted that
their $A_i$ defeats their $B_j$.

(The overall audit is an RLA with risk limit $\alpha$ because the
chance that the overall audit accepts an incorrect outcome is at most
the chance that a particular one of the subaudits with an incorrect hypothesis to
check accepts it as correct.\footnote{Thanks for Philip B. Stark for
  providing this argument.})

\paragraph{Multiple contests}

Multiple contests can be audited concurrently in a similar way.
Each contest is audited using the overall risk limit $\alpha$
as its individual risk limit.  Ballots are sampled uniformly from
the set of all ballots having relevant (still-being-audited) contests
on them.  When a ballot is examined, choices for all relevant (still-being-audited)
contests are determined by hand, and the audits for the relevant
contests make progress (and possibly terminated).  When all audits
have terminated, or all ballots examined for all audits still
being audit, the audit stops.

Each individual contest is audited in a risk-limiting manner.  The
fact that the different audits are using the same ballot samples
doesn't affect this argument (although it may mean that the different
audit results are correlated, which shouldn't be a problem).

\section{Discussion}
\label{sec:discussion}

\CA{} may be easier to explain or understand than Bravo,
and doesn't have any dependence on the unofficial margin.

It would be nice to prove a simple formula for $\beta(n,\alpha)$.
Nonetheless, $\beta(n,\alpha)$ can be adequately estimated
for practical use of \CA{}.

\section{Related Work}
\label{sec:related-work}


DiffSum\cite{Rivest-2015-diffsum}
has an identical structure, but
the constant corresponding $\beta$ there has no formal definition,
and the method is not provably risk-limiting.

We note for the record that there are statistical post-election audits
that don't seem to quite fit the ``risk-limiting audit'' definition,
but which nonetheless have sound probabilistic foundations.  In
particular, the ``Bayesian audit'' of Rivest and
Shen~\cite{Rivest-2012-bayesian} is of this character.  We refer the
reader to that paper for details and discussion.

\section*{Acknowledgments}
\label{sec:acknoledgments}

Ronald L. Rivest gratefully acknowledges support for his work on this
project received from the Center for Science of Information (CSoI), an
NSF Science and Technology Center, under grant agreement CCF-0939370,
and from the Department of Statistics, University of California,
Berkeley, which hosted his sabbatical visit when this work began.
Thanks in particular to Philip Stark for helpful feedback and suggestions.

\bibliography{clip}

\end{document}